\documentclass[conference]{IEEEtran}
\IEEEoverridecommandlockouts
\usepackage{amsfonts}
\usepackage{algorithmic}
\usepackage{algorithm}
\usepackage[dvips]{color}
\usepackage{tabu}
\usepackage{enumitem}
\usepackage{comment}
\usepackage{todonotes}
\usepackage{epsf}
\usepackage{times}
\RequirePackage{scrlfile}
\PreventPackageFromLoading{subfig}
\usepackage{bbold}
\usepackage{mathtools}
\usepackage{mathrsfs}
\usepackage{pdfpages}
\usepackage{epstopdf}
\usepackage{amsmath,epsfig,amssymb,amsthm,cite,url}
\usepackage{float}
\usepackage{graphicx}
\usepackage{textcomp}
\usepackage{geometry}
\usepackage{bbm}
\usepackage{xcolor}
\def\BibTeX{{\rm B\kern-.05em{\sc i\kern-.025em b}\kern-.08em
    T\kern-.1667em\lower.7ex\hbox{E}\kern-.125emX}}
\begin{document}
\title{\huge{Latency and Energy Minimization in NOMA-Assisted MEC Network: A Federated Deep Reinforcement Learning Approach}}
\DeclareRobustCommand*{\IEEEauthorrefmark}[1]{%
  \raisebox{0pt}[0pt][0pt]{\textsuperscript{\footnotesize #1}}%
}
\author{\IEEEauthorblockN{Arian Ahmadi, Anders Høst-Madsen and Zixiang Xiong
\thanks{This research was supported under Grants EECS-1923751 and EECS-1923803.}}
\vspace{-1cm}
\thanks{A. Ahmadi and A. Høst-Madsen are with the Department of Electrical Engineering, University of Hawaii at Manoa, Honolulu, HI 96822 USA (Emails: {aahmadi;ahm}@hawaii.edu).
}
\thanks{Z. Xiong is with the Department of Electrical and Computer Engineering, Texas A\&M University, College Station, TX 77843 USA (Email:
zx@ece.tamu.edu).}
}
\maketitle
\begin{abstract}
Multi-access edge computing (MEC) is seen as a vital component of forthcoming 6G wireless networks, aiming to support emerging applications that demand high service reliability and low latency. However, ensuring the ultra-reliable and low-latency performance of MEC networks poses a significant challenge due to uncertainties associated with wireless links, constraints imposed by communication and computing resources, and the dynamic nature of network traffic. Enabling ultra-reliable and low-latency MEC mandates efficient load balancing jointly with resource allocation. In this paper, we investigate the joint optimization problem of offloading decisions, computation and communication resource allocation to minimize the expected weighted sum of delivery latency and energy consumption in a non-orthogonal multiple access (NOMA)-assisted MEC
network. Given the
formulated problem is a mixed-integer non-linear programming
(MINLP), a new multi-agent federated deep reinforcement learning (FDRL) solution based on double deep Q-network (DDQN) is developed to efficiently
optimize the offloading strategies across the MEC network while accelerating the learning process of the Internet-of-Thing (IoT) devices. Simulation results show
that the proposed FDRL scheme can effectively reduce the weighted
sum of delivery latency and energy consumption of IoT devices in the MEC network and outperform the baseline approaches.
\end{abstract}

\section{Introduction}
The 6G wireless cellular network must support a wide range of new applications with ultra-high reliability and low latency service requirements \cite{semiari2019integrated,ahmadi2021reinforcement}. Among these emerging applications include, connected and autonomous vehicles (CAVs), extended reality (XR), Internet-of-Things (IoT) networks with strict quality-of-service (QoS) requirements on the end-to-end (E2E) latency (e.g., 1 ms) and reliability (e.g., $10^{-8}$ packet loss probability) \cite{3GPP}. Moreover, given the constrained computational power and limited battery capacity of IoT devices, meeting predefined task deadlines that require significant resources often becomes unfeasible for these devices. To meet such stringent service requirements, multi-access edge computing (MEC) is an attractive solution to significantly reduce service latency by enabling base stations (BSs) to process computing tasks (e.g., XR rendering) for user equipment (UE), e.g., IoT devices, directly within the radio access network (RAN) without relying on remote cloud servers \cite{pham2020survey, mao2017survey}.

While promising, ensuring a consistent performance over a MEC network is challenging due to random wireless channel variations, stochastic task arrival, as well as heterogeneity of edge computing servers and computing tasks. Additionally, considering the constrained resources of IoT devices and edge servers, such as limited bandwidth and computing capabilities, the MEC-supported network is susceptible to becoming overwhelmed, resulting in dropped computing tasks and poor QoS. Hence, it is imperative to develop innovative solutions that jointly optimize the offloading decisions and efficiently allocate edge computing resources for processing tasks from IoT devices within the MEC framework.

Recently, the majority of existing research \cite{khalili2019joint, ahmadi2022variational, mao2017survey, pham2020survey, kovalenko2019robust, saleem2020latency} has tackled the challenge of joint decision-making on offloading, communication, and allocation of computational resources in ultra-reliable and low-latency MEC (URLL MEC) utilizing methodologies from the field of optimization theory. The power allocation for delivery latency reduction in a device-to-device (D2D) enabled MEC scenario is studied in \cite{saleem2020latency}. In fact, the authors propose a new power allocation algorithm to minimize the total latency while guaranteeing the task computing
deadline. However, these algorithms have commonly suffered from issues of scalability, time inefficiency, and high computational overhead. Instead of relying on model-based methods, model-free approaches that leverage machine learning (ML) in combination with the computing capabilities of UEs and BSs can provide new opportunities for enabling URLL MEC.

In this regard, the body of work in \cite{mustafa2023reinforcement, li2022federated, xu2022deep, kasgari2020experienced} presents several new schemes based on deep neural networks (DNNs) and deep reinforcement learning (DRL) to optimize the network performance for URLL MEC applications. The authors in \cite{li2022federated} propose a resource allocation technique using multi-agent DRL in a vehicle-to-vehicle communication network. In \cite{xu2022deep}, the authors present a DRL-based computation and communication resource allocation scheme in a MEC railway IoT network. In \cite{wang2020machine}, the authors develop a joint task, spectrum and
transmit power allocation method based on multi-stack RL to minimize the computational
and transmission latency in a MEC network. However, a significant drawback of many DRL algorithms is their centralization, leading to scalability issues as the number of devices increases. Additionally, finding an optimal policy becomes computationally complex, especially with the exponential growth of the state and action spaces. Moreover, centralized learning demands IoT devices to share their data to train a global model, potentially compromising privacy.

Recently, distributed learning algorithms that enable users to collaboratively build a unified learning model while conducting local training \cite{park2019wireless} are developed. One of the most promising distributed learning frameworks is federated
learning (FL) which preserves the data privacy by avoiding data uploading to the parameter server (PS) \cite{mcmahan2021advances,bonawitz2019towards}. In FL scheme, each device utilizes its individual local dataset for training and subsequently transfers its respective local model to the PS for global aggregation. In particular, FL improves collaboration among agents and enhances the scalability of network resource management algorithms. In this regard, the problem of joint power allocation and
resource allocation for URLL communications in vehicular networks is investigated in \cite{samarakoon2019distributed}. The authors propose a novel distributed scheme
based on FL in order to estimate the tail
distribution of the queue lengths. However, the constraint of energy consumption of each user in the network is not considered.

More recently, several research studies have delved into the challenge of reducing latency and energy consumption for IoT devices within the context of a FL system \cite{chen2022federated,pei2023federated, mills2019communication}. For instance, in \cite{chen2022federated}, the authors investigate the problem of jointly
optimizing resource and learning performance to reduce
communication costs and improve learning performance in
wireless FL systems. The authors in \cite{mills2019communication} focus on optimizing various aspects of FL, including weight compression, convergence analysis and iteration reduction over IoT networks. However, none of the mentioned research works applied FL to enhance the effectiveness in solving a real-world wireless resource allocation problem. In \cite{han2019federated}, the authors employ FL to facilitate the learning process in DRL, wherein individual local DRL models were trained and subsequently merged collaboratively to construct a comprehensive global DRL model. However, the authors modeled the network as a queuing system without explicitly ensuring QoS for users' tasks and allocation of computation resources.

The main contribution of this paper is a novel framework to optimize the reliability of the MEC IoT network by minimizing joint delivery latency and energy consumption of IoT devices using federated DRL (FDRL). In particular, we design a non-orthogonal multiple access (NOMA) model of heterogeneous network
with multiple BSs and multiple IoT devices to dynamically optimize offloading decisions, base station assignment, and channel resource allocation in a decentralized manner across the RAN, while accounting for the constraints of the wireless network and edge computing servers. Given that the proposed problem is a mixed integer non-linear programming (MINLP) and difficult to solve, a new algorithm is developed to jointly solve the offloading decision problem, computing resource and transmit power allocation across the MEC  network. Mainly, we first reformulate our problem as a multi-agent DRL problem and solve
it using double deep Q-network (DDQN). To enhance both the quality and speed of learning of the proposed algorithm, we integrate FL at the end of each episode. Utilizing FL results in a framework that is both privacy-preserving and scalable, establishing a context for cooperation among agents. Extensive simulation results are carried out to show
the superiority of the proposed algorithm through comparisons with those existing schemes. Results indicate that the proposed scheme considerably
outperforms the benchmarks under multiple performance metrics.

The rest of the paper is organized as follows. Section \ref{System Model} presents the system model. Section \ref{MULTI-OBJECTIVE PROBLEM FORMULATION} and \ref{PROPOSED FEDERATED DDQN ALGORITHM} describe the problem formulation and the proposed solution. Simulation results are provided in Section \ref{Simulation Results} and conclusions are presented in Section \ref{Conclusion}.
\vspace{-0.15cm}
\section{System Model}
\label{System Model}
We consider an MEC  heterogeneous network based on NOMA consisting of a set $\mathcal{M}$ of $M$ IoT devices with limited computation and energy resources and a set $\mathcal{N}$ of $N$ BSs as shown in Fig. \ref{fig:System Model}. At each time $t$, device $m$ needs to process one of the tasks in its queue and can either execute its task
locally or offload it to a nearby BS $n$. Specifically, we use
$x_{m,n} \in \{0, 1\}$ to indicate the IoT device offloading decisions. When
$x_{m,n} = 0$, represents the IoT device uses local processing, Otherwise, the
it uploads to BS for processing. Meanwhile, NOMA transmission scheme applies successive interference cancellation (SIC) receivers at the receiving end to realize multi-user detection. Next, we explain the transmission and
computing latencies in details.
\subsection{Over-the-Air Transmission Latency and Energy Consumption}
If the IoT device $m$ decides to offload its task, it should first transmit it to the BS through wireless channels. 
Due to the wireless fading channel, some packets may not be decoded successfully at the BS, hence, re-transmission is needed. With this in mind, the transmission latency is given by: 
\begin{equation}\label{uplink latency}
\tau^{air}_{m,n}= \sum_{j =1}^{J}\left\lceil\frac{I_{m}}{r_{m,n,j}}\right\rceil,\vspace{-0.15cm}
\end{equation}
where $\lceil.\rceil$ is the ceiling function, $J-1$ is the number of re-transmissions, and $I_{m}$ is the size
of the task (in bits). In (\ref{uplink latency}), $r_{m,n,j}$ represents the transmission data rate for the $j$-th transmission is calculated as follows:
\begin{equation}
r_{m,n,j}=\omega \log _{2}\left(1+\gamma_{m,n,j}\right),\vspace{-0.15cm}
\end{equation}
where $\omega$ is the channel bandwidth. Moreover, $\gamma_{m,n,j}$ is the signal-to-interference-plus-noise-ratio (SINR) and is given by
\begin{equation}\label{SINR_uplink}
\gamma_{m,n,j}=\frac{G_{m} G_{n} P_{m} h_{m,n,j} L_{m,n}}{\sum_{m^{\prime} \neq m} P_{m^{\prime},n}+\sigma_{n}^{2}},\vspace{-0.15cm}
\end{equation}
where $P_m$, $P_{m^{\prime},n}$, and $\sigma_{n}^{2}$ denote, respectively, the transmit power of IoT device $m$, the received power from an interfering IoT device $m^{\prime}$, and the noise power. $G_m$ and $G_n$ are the antenna gains for IoT device $m$ and BS $n$, respectively. In addition, $h_{m,n,j}$, and $L_{m,n}$ represent, respectively, the Rayleigh fading channel gain for the $j$-th transmission and path loss between IoT device $m$ and BS $n$. The channel gain $h_{m,n,j}$ is considered flat-fading over the bandwidth $\omega$ and constant during the transmission of one packet. 

In addition, energy consumption of device
$m$, while offloading its task to the BS $n$ is given as follows
\begin{equation}
\begin{aligned}
E^{air}_{m,n}&= \tau^{air}_{m,n}P_{m}
\\&=\frac{I_{m}P_{m}}{\omega \log _2\left(1+\frac{G_{m} G_{n} P_{m} h_{m,n,j} L_{m,n}}{\sum_{m^{\prime} \neq m} P_{m^{\prime},n}+\sigma_{n}^{2}}\right)}.
\end{aligned}
\end{equation}

\subsection{Computing Latency at Edge Computing Servers }\label{Computational}
Computing latency refers to the time needed for executing a task at an edge computing server within the MEC network. The execution time of a task depends on the data to be processed and the tasks submitted, therefore, it can be modeled as a random variable \cite{tareq2018ultra}. For instance, the execution time for performing the object detection highly depends on the quality or level of details in the captured images, as well as the type (GPU vs. CPU) and processing resources (e.g., processing bandwidth) of the edge server.

Let $f_{max}$ denote the maximum computing-cycle frequency of edge processors for each BS. If device $m$ offloads its task to the edge server the computation
latency would be $\tau^{comp}_{m,n} = \frac{c_{m}}{f^{edge}_{m,n}}$, where ${c_{m}}$ and ${f^{edge}_{m,n}}$ denote the CPU cycle requirement
of the task and average
computation capacity of edge server, respectively. From the standpoint of an IoT device, the energy expended to process a task when offloaded to an edge server includes the energy used for task transfer. Hence, the overall energy utilization for edge computing would be equal to $E_{m,n}^{air}$.

In addition, if device $m$ decides to process its task locally, the latency and energy consumption for local computation would vary based on the computation resources allocated for task processing at time $t$, denoted by $f^{loc}_{m}$. Therefore, the local latency and energy consumption for the device are modeled as follows:
\begin{equation}
\begin{aligned} 
&\tau^{loc}_{m}=\frac{c_{m}}{f^{loc}_{m}}, \\&E^{loc}_{m}=\zeta_m\left(f^{loc}_{m}\right)^2,
\end{aligned}
\end{equation}
where $\zeta$ is a constant coefficient which depends on the chip
architecture in devices. It's important to note that higher resource utilization, whether in terms of transmit power or computation capacity, reduces the delivery latency at the cost of higher energy consumption. Hence, managing this trade-off requires careful handling through efficient offloading decision-making and precise optimization of both local computation and offloading strategies.
\begin{figure}[t!]
	\centering
	\centerline{\includegraphics[width=8cm,height=6cm]{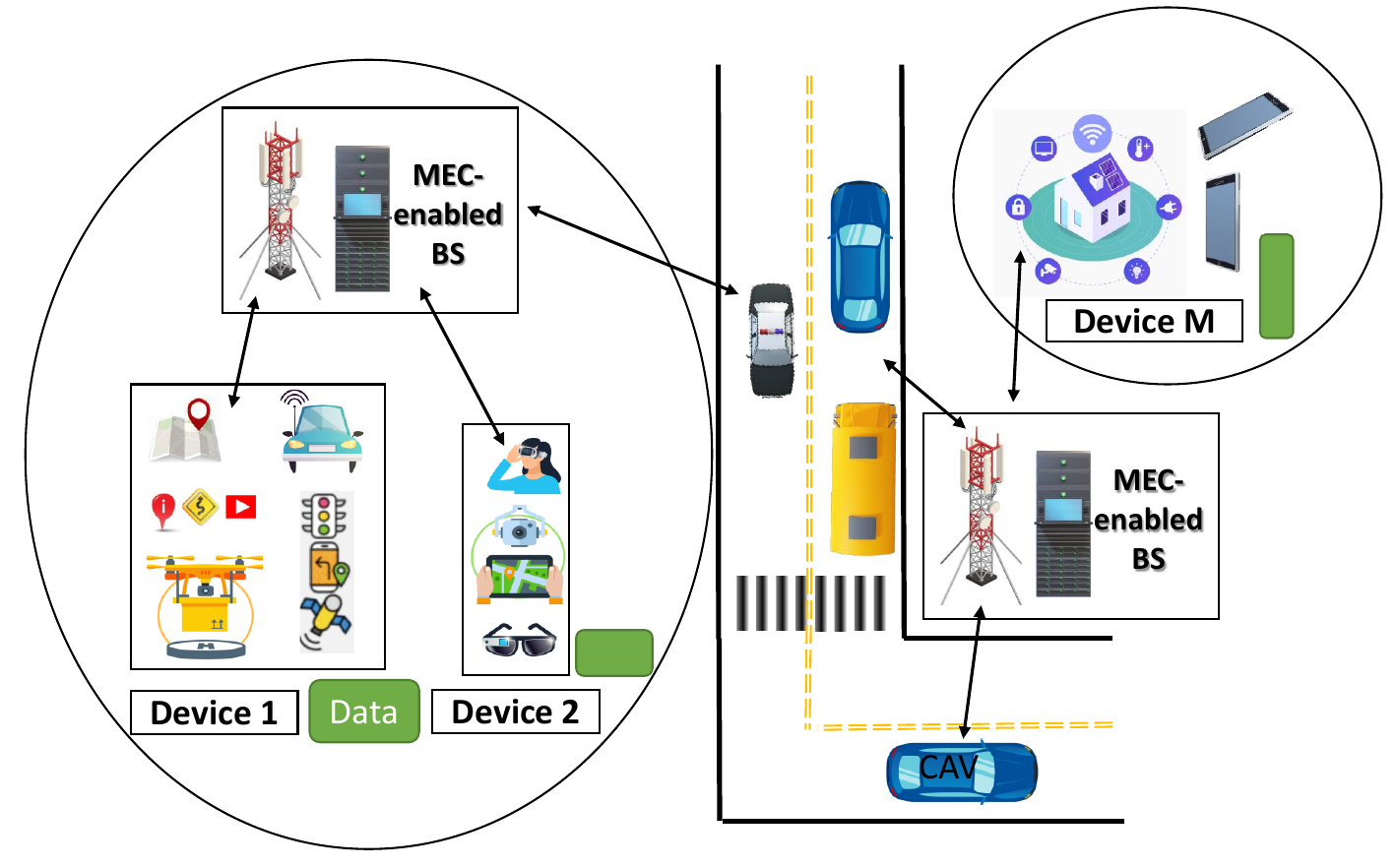}}\vspace{-0.6em}
	\caption{\small System Model.}\vspace{-1em}
	\label{fig:System Model}
\end{figure}
\section{MULTI-OBJECTIVE PROBLEM FORMULATION}
\label{MULTI-OBJECTIVE PROBLEM FORMULATION}
The main goal of this paper is to minimize the latency of completing task and
energy consumption at the same time by jointly accounting the offloading decisions, BS selection and channel resource allocation made by IoT devices. The long-term expected weighted sum of latency and energy consumption between each IoT device $m$ and BS $n$, is formulated as follows:
\begin{equation}
\begin{aligned}
& \tau_{m,n}\left(\mathbf{P}, \mathbf{f}^{loc}, \mathbf{f}^{edge},\mathbf{x}\right)=\\
& \mathbb{E} 
\left[ \lim_{t\to\infty} \frac { 1 } { t } \sum _ { i = 0 } ^ { t } (1-x_{m,n,i}) \tau^{loc}_{m,i}+x_{m,n,i}(\tau^{comp}_{m,n,i}+\tau^{air}_{m,n,i})\right],
\end{aligned}
\end{equation}
and
\begin{equation}
\begin{aligned}
&E_{m,n}\left(\mathbf{P}, \mathbf{f}^{loc}, \mathbf{f}^{edge}, \mathbf{x}\right)=\\
& \mathbb{E}\left[\lim _{t \rightarrow \infty} \frac{1}{t} \sum_{i=0}^t(1-x_{m,n,i}) E^{loc}_{m,i}+x_{m,n,i} E_{m,n,i}^{air}\right],
\end{aligned}
\end{equation}
where $\mathbf{P}$, $\mathbf{f}^{loc}$, $\mathbf{f}^{edge}$ and
$\mathbf{x}$ represent the vectors of transmit
powers, local computing resource allocation, edge server computation resource allocation, local computing and
edge offloading decision,
respectively. 

For any given device $m$ at time $t$, we model the decision-making problem (DMP), expressed as:
\begin{equation}\label{Optimization}
\begin{aligned}
& \quad \min _{\mathbf{P}, \mathbf{f}^{loc}, \mathbf{f}^{edge},\mathbf{x}} U =  \tau_{m,n}+\lambda E_{m,n} \\
& \quad \text { subject to: } \\
&\mathrm{C} 1: (1-x_{m,n}) E^{loc}_{m}+x_{m,n} E^{air}_{m,n}\leq E_{\max}, \\&
\mathrm{C} 2: (1-x_{m,n})\tau^{loc}_{m}+x_{m,n} (\tau^{air}_{m,n}+\tau^{comp}_{m,n})\leq \tau_{\text{th}},
\\& \mathrm{C}3: f^{loc}_{m}\leq F_{\max},  \\& 
\mathrm{C4:} \sum_{m \in \mathcal{M}} f^{edge}_{m,n} \leq f_{\max }, \\
&
\mathrm{C} 5: x_{m,n}  \in\{0,1\}, \quad \quad \quad \quad \quad \forall m\in \mathcal{M}, \forall n \in \mathcal{N},
\end{aligned} 
\end{equation}
where  $\lambda$ is a weighting factor. Constraint $\mathrm{C1}$ represents the
restriction on the energy resource of the device and that energy
utilization should not exceed $E_{max}$.
Constraint $\mathrm{C2}$ expresses that total time for processing the task  must meet the maximum latency threshold determined by QoS requirement regardless of whether it is following local processing or offloading to
BS processing. Constraints $\mathrm{C3}$ and $\mathrm{C4}$ indicate the local computation capacity with
the maximum threshold $F_{max}$ and the sum of the computation
frequency allocated to IoT devices should not exceed the maximum
computation frequency of the MEC server $f_{\max}$.  The constraints $\mathrm{C5}$ indicates that the decision
variables of the problem are all binary indicators.

The proposed optimization problem in (\ref{Optimization}) is a MINLP, hence, it is difficult to solve. Next, we develop
a new efficient algorithm to solve this problem.

\section{PROPOSED FEDERATED DDQN ALGORITHM}
\label{PROPOSED FEDERATED DDQN ALGORITHM}
When addressing the joint optimization problem in multi-user scenarios, various challenges come to light.
\begin{algorithm}[ht]
	\caption{Proposed FDRL for Joint Offloading Decision and Computing and Communication Resource Allocation (Training Phase)}
	\label{alg:exmp}
	\renewcommand{\algorithmicrequire}{\textbf{Input:}}
	\renewcommand{\algorithmicensure}{\textbf{Output:}}
	\begin{algorithmic}[1]
		\REQUIRE $\mathcal{N}$, $\mathcal{M}$, $t=0$ 
		\ENSURE $\mathbf{P}, \mathbf{f}^{loc}, \mathbf{f}^{edge},\mathbf{x}$ \\
         \STATE Initialize the global model $\phi^{\text {global}}$, the online and target networks for each agent $m$.
		\WHILE{(maximum number of iterations is not reached)}
		\STATE $\phi_{m}^{\text{online}}=\phi_{m}^{\text{target}}=\phi_{m}^{\text{global}}$,
		\STATE Using (\ref{DeviceSelection}), choose the set of participating devices, $\mathcal{K}$.
        \FOR{each device $m$ in $\mathcal{K}$}
            \FOR{each time step $t$ if $|\mathcal{I}_{m}| > 0$}
            \STATE Compute offloading decision, $\mathbf{x}$, using $\phi_{m}^{\text{online}}$,
            \STATE Interact with environment and solve  (\ref{localOPT}) or (\ref{edgeOPT}) applying KKT conditions,
            \STATE Save the experience $\mathbf{P}$, $\mathbf{f}^{loc}$ and  $\mathbf{f}^{edge}$ in replay memory $\mathcal{O}_{m,n,t}$,
            \STATE Train the local model on $\mathcal{O}_{m,n,t}$,
            \STATE Transmit  $\phi_{m}^{\text{online}}$ to the PS,
            \ENDFOR
        \ENDFOR
        \STATE  update $\phi_{m}^{\text{global}}$ using (\ref{UpdateGlobal}) and distribute it  to all selected devices in the next round.
	\ENDWHILE 
	\end{algorithmic}
\end{algorithm}

\begin{itemize}
    \item  The significant mobility observed in IoT devices results in frequent and unpredictable shifts within the communication channel. This dynamic variability poses a considerable challenge for IoT devices, impeding their ability to effectively make optimal real-time decisions.
\end{itemize}

\begin{itemize}
    \item Due to constrained resources, the decision made by each individual IoT device holds a consequential influence on the selection and actions of other devices within the network, meaning that optimal decision-making by one device not only affects its own resource allocation and performance but also ripples through the network, influencing the decisions and performance of other devices.
\end{itemize}

\begin{itemize}
    \item As the number of IoT devices grows in the network, the offloading decision complexity increases exponentially. This escalating complexity underscores the need for scalable and efficient approaches to address the evolving demands of IoT networks. 
\end{itemize}

To address the issues mentioned above, we propose a multi-agent DDQN algorithm to solve the secure offloading and computing and communication resource allocation problem. The details of the algorithm are
elaborated in the following section.

\subsection{Sketch of double deep Q-network}
Given the aforementioned challenges, employing traditional optimization methods to address the dynamic optimization problem described in equation (\ref{Optimization}) is not feasible. Model-free DRL is a useful tool for handling the DMP and learning the optimal
solutions in dynamic environments. Hence, the DMP is formulated as a Markov Decision Process (MDP). Particularly, we model our problem as a multi-agent DDQN problem. For each device (DRL agent),
we have following components:
\begin{itemize}
    \item \textit{State space:} the state space for each agent $m$, denoted by $s_m$, is defined as \\ $s_m=\left\{\mathcal{I}_{m}, h_{m,n}, I_{m}, c_{m}, E_{\max}, F_{\max}\right\}$, where $\mathcal{I}_{m}$ is the
length of the task queue of device $m$.
    \item \textit{Action space:} The action space of agents, denoted by $\mathcal{A}$, includes the  offloading decisions depending on the current state.
    \item \textit{Cost function:} The objective function defined in (\ref{Optimization}) depends on the value of $P_{m}$ and $f_{m,n}^{edge}$ when offloading to edge server and $f_{m}^{loc}$ if local computation is selected. Hence, in order to accurately capture the benefits of a specific offloading decision within the cost function, it is imperative to carefully optimize these variables. Therefore, when device $m$ performs its task locally, i.e., $x_{m,n} = 0$, the cost would be calculated by solving the following optimization problem:
\begin{equation}\label{localOPT}
\begin{gathered}
\min _{f_{m}^{loc}} \tau_{m}^{loc}+\lambda E_{m}^{loc} \\
\text { subject to: C1, C2, C3. }
\end{gathered}
\end{equation}
If device $m$ offloads its task to the BS, i.e., $x_{m,n} = 1$, the transmit power and computing frequency at the edge server would be optimized by solving the following optimization problem:
\begin{equation}\label{edgeOPT}
\begin{aligned}
& \min _{P_{m}, f_{m,n}^{edge}} (\tau_{m,n}^{comp}+\tau_{m,n}^{air})+\lambda E_{m,n}^{air}  \\
& \text { subject to: } \mathrm{C} 1, \mathrm{C} 2, \mathrm{C} 4.
\end{aligned}
\end{equation}
\end{itemize}
It's important to note that both equations (\ref{localOPT}) and (\ref{edgeOPT}) represent convex optimization problems with respect to the variables $f_{m}^{loc}$ and $P_{m}$, $f_{m,n}^{edge}$, respectively. These types of optimization problems can be effectively solved using standard software tools, e.g., Karush-Kuhn-Tucker (KKT) conditions. After finding the optimal computing and communication resource allocation vectors, $\mathbf{f}^{loc}$, $\mathbf{P}$ and $\mathbf{f}^{edge}$, we feed
them into the DDQN framework as the immediate cost function. After the DDQN agent is trained through this process for one training round, we apply a FL framework where each IoT device will train
its DDQN models, share and update their models with PS. 
\begin{table}[t!]
	\footnotesize
	\centering
	\caption{\vspace*{-0cm}  Simulation Parameters}\vspace*{-0.2cm}
	\begin{tabular}{|>{\centering\arraybackslash}m{0.9cm}|>{\centering\arraybackslash}m{3.7cm}|>{\centering\arraybackslash}m{2.5cm}|}
		\hline
		\bf{Notation} & \bf{Parameter} & \bf{Value} \\
		\hline
		$N$    & Number of BSs &  5\\
		\hline
		$M$ & Number of IoT devices &  30 \\
	    \hline
		$P_m$    & Transmit power of a IoT device & 10 mW\\
		\hline
		$G_n$,$G_m$ & Antenna gains & 1\\
		\hline
		$f_{max}$   & Computing frequency & 30 GHz \\
		\hline
		$N_0$ &  Noise power spectral density & $-90$ dBm/Hz \\
		\hline
		$\omega$ &  Total system bandwidth & 50 MHz \\
		\hline
  $\tau_{\text{th}}$ & Service latency requirement& 10-100 ms\cite{semiari2019integrated}\\
		\hline
	\end{tabular}\label{tabsim1}\vspace{-0.35cm}
\end{table} 
\subsection{DDQN training phase}
Consider the immediate cost of each IoT device $m$ obtained
from the solution of the (\ref{localOPT}) and (\ref{edgeOPT}) as $cost_m(s,a)$. Using Bellman equation, the action-state value is:
\begin{equation}
Q_m(s, a)=cost_m(s, a)+\gamma \sum_{s^{\prime} \in \mathcal{S}} W_{s s^{\prime}}(a) \max _{a^{\prime} \in \mathcal{A}} Q^{*}_m\left(s^{\prime}, a^{\prime}\right),
\end{equation}
where $\mathcal{S}, W_{s s^{\prime}}(a)$, and $0<\gamma<1$ are the set of states, the transition probability function, and the discount factor, respectively. Our goal by using DDQN is
to find a solution to minimize the state-action function $Q^{*}_m\left(s^{\prime}, a^{\prime}\right)$. Each agent $m$ has two neural networks working alongside each other, one called \textit{online network} with parameters $\phi_m^{\text {online }}$ and the other called \textit{target network} with parameters $\phi_m^{\text {target}}$. At each training iteration the target value for training the online
network in device $m$ is calculated as:
\begin{equation}
\begin{aligned}
Z_m=cost_m(s,a)+\gamma Q_m(s^{\prime},\max_{a^{\prime} \in \mathcal{A}}
Q^{*}_m(s^{\prime},a^{\prime} ; \phi_m^{\text{online}}), \phi_m^{\text{target}})
\end{aligned}
\end{equation}
To overcome related issues regarding to train a DRL agent in a centralized manner such as scalability and privacy and to improve the overall performance gains, we propose FDRL that empowers each agent to train its own local model,
using its own local data. Then these local models are sent to a
PS to be combined together. This process
continues until a certain criterion is met.
\subsection{Federated DRL Scheme}
Training FDRL agents is completed in three steps. First, small
subset of $M$ IoT devices, denoted by $\mathcal{K} = {1, ..., K}$ is selected to contribute in
FL based on the
following criterion:
\begin{equation}\label{DeviceSelection}
\max _{m \in \mathcal{M}} \operatorname{Variance}\left(\frac{d_m P_{\max}}{F_{\max}}\right),
\end{equation}
where $d_m$ represents the distance of device from BS. Then, each selected IoT
devices use DDQN to train their local models and send the weights of online network, $\phi_m^{\text {online}}$, to the PS. Finally, the PS aggregates the
models using FedAvg \cite{mcmahan2017communication} and forms a single global model that would be
then transmitted to all selected devices. The model aggregation is given as:
\begin{equation}\label{UpdateGlobal}
\phi^{\text {global}}= \frac{\sum_{m \in \mathcal{K}} \phi_m^{\text{online}}}{|\mathcal{K}|}
\end{equation}
The proposed algorithm is summarized in Algorithm \ref{alg:exmp}.  

\section{Simulation Results}
\label{Simulation Results}
In this section, we evaluate the performance of the proposed scheme for reliability optimization in the MEC network shown in Fig. \ref{fig:System Model}. The packet size of each IoT devices is selected randomly from a uniform distribution with a range
[1, 10] kbits. Simulation parameters are summarized in Table \ref{tabsim1}. We compare the performance of the proposed method with two
baseline approaches. The first baseline approach, hereinafter
referred to as “Baseline 1”, uses simple distributed DDQN with no aggregation (non-federated) \cite{qi2021federated} to solve the proposed problem. The second baseline, hereinafter referred to as “Baseline 2”, is the state-of-art centralized deep Q-network (DQN) \cite{giannopoulos2021deep} scheme where there is a centralized entity that makes decisions based on the collective information from all agents. The performance was evaluated by averaging the results over sufﬁciently large Monte Carlo runs. 

\begin{figure}[t!]
	\centering\centerline{\includegraphics[width=8cm]{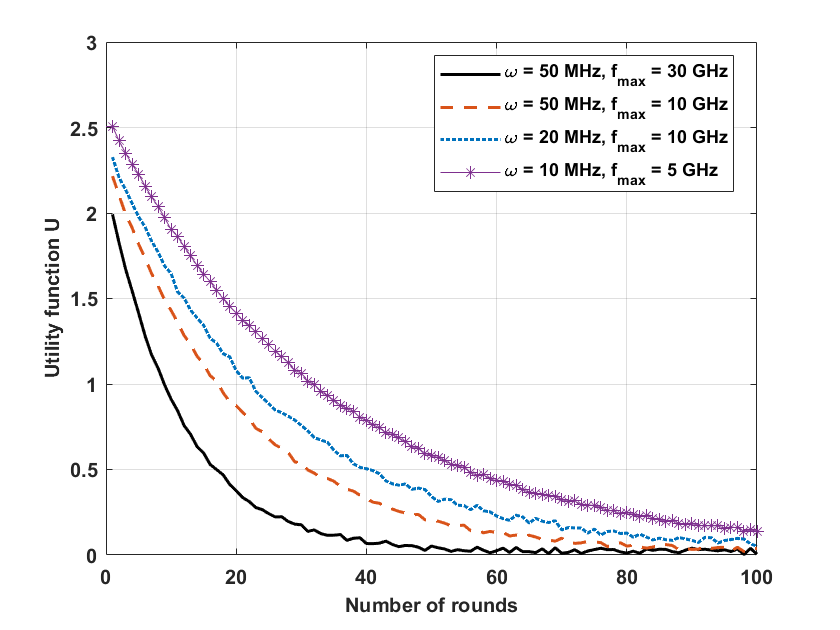}}\vspace{-0.5em}
	\caption{\small The convergence property of the FDRL algorithm.}\vspace{-1.2em}
	\label{LOSS}
\end{figure}

\begin{figure}[t!]
	\centering	\centerline{\includegraphics[width=8cm]{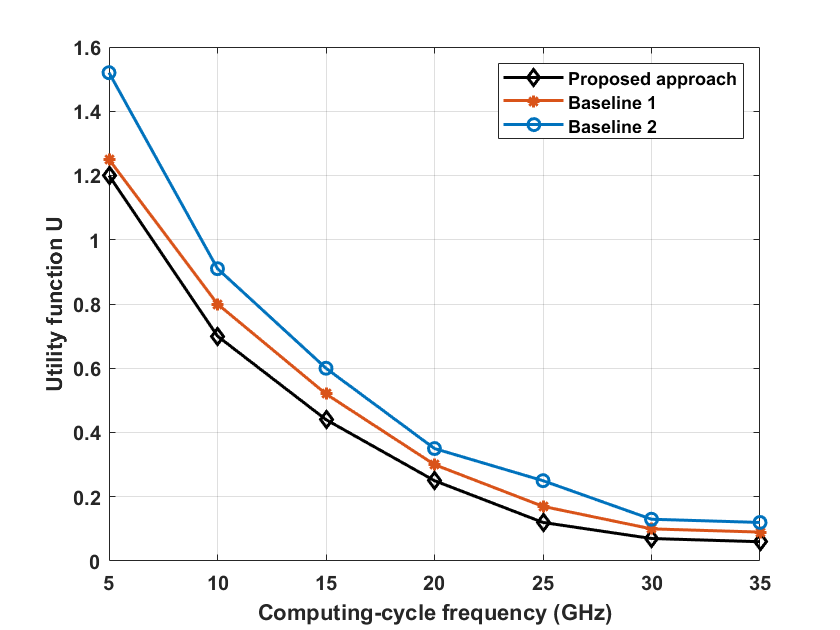}}\vspace{-1em}
	\caption{\small Utility function versus the computing frequency.}\vspace{-1em}
	\label{COMFREQ}
\end{figure}

\begin{figure}[t!]
	\centering
	\centerline{\includegraphics[width=8cm]{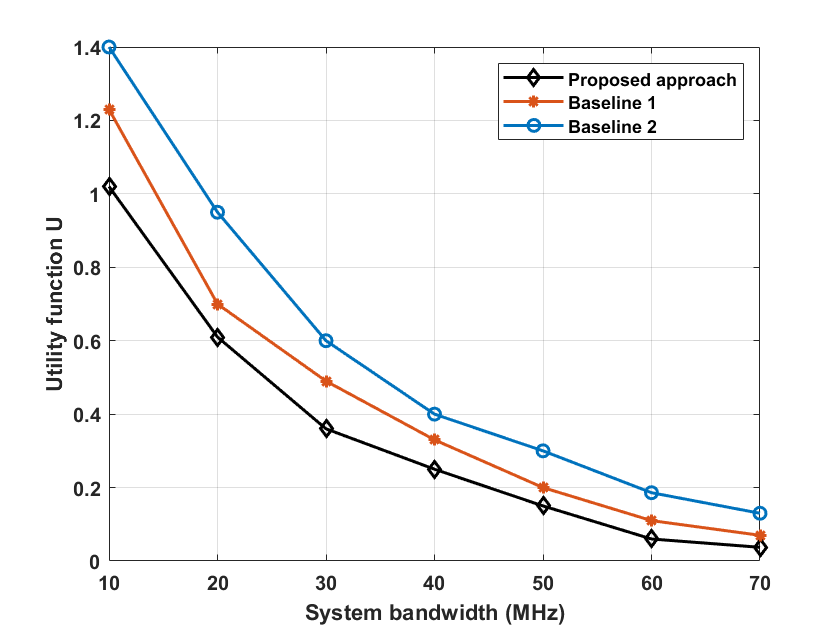}}\vspace{-0.6em}
	\caption{\small  Utility function versus the bandwidths.}\vspace{-1em}
	\label{power}
\end{figure}

\begin{figure}[t!]
	\centering\centerline{\includegraphics[width=8cm]{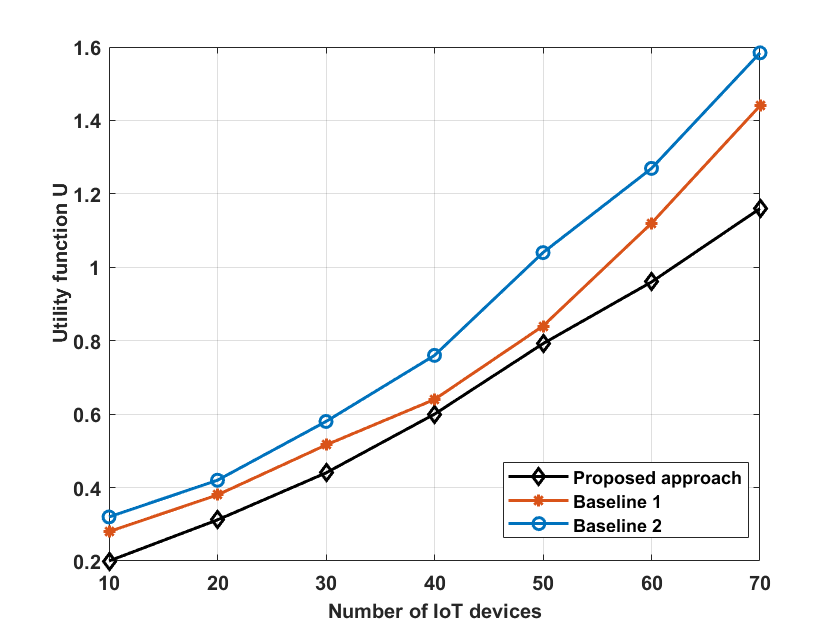}}\vspace{-1em}
	\caption{\small Utility function versus the number of IoT devices.}\vspace{-1.5em}
	\label{Hist}
\end{figure}

Figure. \ref{LOSS} shows the convergence of the proposed FDRL algorithm versus the number of epochs for different system parameters. From Fig. \ref{LOSS}, we observe that the proposed
algorithm has a fast convergence rate, and can converge
within reasonably small number of epochs for all the simulated cases. 

Figure. \ref{COMFREQ} compares the utility function for the proposed approach with the baseline methods, versus the computing-cycle frequency of each MEC. Simulation results show that the utility function
declines as the computing capacity increases for all three methods. The results in Fig. \ref{COMFREQ} also indicate that the proposed scheme
surpasses the other two methods. For example, for $f^{edge}$ = 25 GHz , the performance gain is up to 24\% and 38\%, respectively,
compared to baseline schemes 1 and 2 when $N$ = 5 BSs and
$M$ = 30 IoT devices.

Figure. \ref{power} illustrates how the utility function is affected by the allocation of different bandwidth. As the bandwidth increases, the utility function decreases due to the fact that each IoT device requires to dedicate small amount of power to achieve higher transmission rates. Therefore, both delivery latency and energy consumption are reduced. The results in Fig. \ref{power} indicate the superior performance of the presented scheme compared to the baseline methods. For instance, in a MEC network with assigned bandwidth $\omega$ = 40 MHz, the performance gains yielded by the proposed algorithm are up to 15\% and 27\%, respectively, compared to baseline methods 1 and 2.

In Fig. \ref{Hist}, the utility function versus the network size is shown for the proposed approach and the
two baseline methods. It is clear that the utility function
increases as more IoT devices exist in the network. The results in
Fig. \ref{Hist} highlights that the proposed
scheme can yield up to 13\% and 30\% performance gain, when $M$ = 40, compared to the baseline 1 and 2, respectively. Furthermore, Fig. \ref{Hist} also shows the scalability of the proposed method. For example, with the utility function of 0.9, the proposed scheme can support up to 60 IoT devices, which is 18\% and 33\% higher compared to baselines 1 and 2, respectively.

\section{Conclusions}
\label{Conclusion}
In this paper, we addressed a joint latency and energy minimization problem for the NOMA-assisted MEC IoT network to guarantee the reliability of the network. Considering the strict QoS requirement for the IoT devices, the problem was formulated by jointly optimizing the offloading decisions, computing recourse allocation and transmit power control. To solve the proposed MINLP, we have developed a new algorithm that adopted FL, DDQN, and optimization theory called FDRL. More specifically, DDQN is used to determine the offloading decisions of the IoT devices. Given the offloading decisions, the computing capacity or transmit power of the devices is optimized to minimize the utility function. Then, we feed
the results into the DDQN framework as the immediate cost function to optimize the offloading decisions. To improve the learning speed of IoT
devices, we use FL at the end of each round.
Simulation
results have confirmed the effectiveness of the proposed scheme to those comparative algorithms.

\bibliographystyle{IEEEbib}
\bibliography{references}
\end{document}